\documentclass[showpacs,amsmath,amssymb,twocolumn,floatfix,pra]{revtex4}
\usepackage[dvips]{graphicx}

\def\na{n_\textrm{A}}
\def\nb{n_\textrm{B}}
\def\tmax{t_{\textrm{max}}}
\def\nmax{N_{\textrm{max}}}
\def\ket#1{|#1\rangle}

\def\braket#1#2{\langle #1 | #2 \rangle}

\begin{document}
\title{Quantum computing of semiclassical formulas}

\author{B.~Georgeot and O.~Giraud}

\affiliation{Laboratoire de Physique Th\'eorique,
Universit\'e de Toulouse, CNRS, 31062 Toulouse, France}

\date{January 30, 2008}

\begin{abstract}
We show that semiclassical formulas such as the Gutzwiller trace
formula can be implemented on a quantum computer more efficiently
than on a classical device.  We give explicit quantum algorithms
which yield quantum observables from classical trajectories,
and which alternatively test the semiclassical approximation
by computing classical actions from quantum evolution.
The gain over classical computation is in general quadratic,
and can be larger in some specific cases.

\end{abstract}
\maketitle


\section{introduction}
It is now widely recognized that the principles of quantum mechanics 
allow to realize new computational devices which can be more efficient
than their classical counterparts \cite{feynman,josza,steane,nielsen}.
Quantum algorithms have been proposed which take advantage of the quantum
mechanical properties of these devices to perform specific tasks 
faster than on a classical computer.  The most famous such
algorithm is due to Shor \cite{shor} and factors large integers exponentially
faster than any known classical algorithm. Another algorithm,
for which the gain is only quadratic, enables to search
an unsorted database \cite{grover}.
Efforts have been devoted also to using such quantum computers to simulate 
the behavior of complex 
physical systems, a task of much practical interest.  Algorithms have been set
up enabling to simulate certain quantum mechanical systems efficiently
\cite{lloyd,BG1,ablloyd,saw1,interm}, as was
originally envisioned by Feynman.  
However, as quantum algorithms use procedures different from 
classical algorithms, it is by no means obvious which problems
can be sped up by using a quantum computer.  It is therefore important 
to precisely specify the class of problems that can be solved efficiently
on a quantum computer, especially among problems which have
 been implemented by scientists on classical devices because
of their practical interest.

On classical computers, a great deal
of activity in the past decades
has been devoted to the numerical implementation of 
{\em semiclassical formulas}.
Such formulas approximate quantum mechanics
through classical quantities, and 
have been used since the beginning of quantum mechanics.  Although
they have been much studied, their application 
to practical computation of quantum observables
 is often hampered by the exponential proliferation of 
classical orbits involved when the system is chaotic.  Semiclassical
formulas enable
to approximate the exact quantum mechanics for small $\hbar$, and give an
insight into the relationship between classical and quantum mechanics.
For integrable systems with $n$ degrees of freedom, 
classical dynamics takes place on $n$-dimensional tori in the $2n$-dimensional 
phase space.  
In this case, semiclassical formulas quantize individual tori. They
are relatively
straightforward to implement and have been constructed and used early
in the development of quantum mechanics.  In contrast,
for chaotic systems this quantization of tori
is not valid, as pointed out by Einstein as early as in 
1917 \cite{einstein}, and
individual wavefunctions cannot be built from a single classical structure.
As a substitute,
 various formulas have been constructed, which express the
quantum quantities in terms of an (infinite) set of classical orbits.
The most famous such formula is the {\em Gutzwiller trace formula}
\cite{gutz}, where
the quantum density of states $d(E)=\sum_n \delta (E-E_n)$ 
(where $E_n$ are the energy levels) is written as a function of all 
classical periodic orbits of the
system. It has the general form  $d(E)\equiv \sum_p A_p e^{i\varphi_p/\hbar}$, 
where the sum runs over all
periodic orbits, $A_p$ is related to the stability
of the orbit and $\varphi_p$ to its action.
It can be viewed as a Fourier-type duality between the set of all
eigenenergies of the system on the one hand and the set of all actions
of periodic orbits on the other hand.  Other formulas of the same kind
give the quantum propagator $G(x,x')$ in terms of all
classical orbits from $x$ to $x'$ (Van Vleck formula) \cite{vanvleck}
or scattering amplitudes
in term of scattering orbits (Miller's formula) \cite{miller}.  
Many works
have implemented numerically such formulas by truncating the sum over
classical orbits
(see e.g. \cite{cnc,berry,gregor,charles,uzy,daniel}), e.g. to obtain
the semiclassical spectrum, but because of the exponential
proliferation of classical orbits typical of chaotic systems
only a few semiclassical eigenvalues can be extracted.
Several methods have been devised to reduce the number of orbits entering
the sum \cite{curvature,berrykeating,harmonic}, but they all require
summing up contributions from a still exponential number of orbits.

In this paper, we study the implementation of
semiclassical formulas on quantum computers. We 
show that for certain dynamical systems, 
such formulas can be computed more efficiently on a quantum
computer than on a classical device.  
From the quantum information point of view,
this gives new examples of algorithms where a gain can
be reached compared to classical algorithms. 
From the 
point of view of quantum chaos, this would enable these formula to become
more practical on a quantum computer if such a device
becomes available, and thus to explore
the quantum-classical correspondence in regimes 
which are difficult to reach on a
classical computer. 
The paper is organized as follows. In section II, we present 
in detail the most famous semiclassical formula
which relates the density of states to classical periodic orbits
(Gutzwiller trace formula), in the specific case of quantum maps.
We then discuss in section III a quantum algorithm 
which implements this semiclassical formula in the form where it is
most difficult classically,  i.e. summing up classical orbits and extracting 
quantum observables.  
In section IV we implement the same formula but in 
the reverse direction,
i.e. using quantum observables to extract classical quantities.  
Our method can be considered as a new 
way of extracting information from the 
quantum simulation of quantum systems.
Indeed, 
while many quantum systems can be
simulated efficiently on a quantum computer, a crucial point to get a complete
algorithm and make the gain effective
is to devise a readout method once the simulation is performed.
It has been shown that the gain over classical computation
can depend critically
on the observable measured at the end of the simulation 
\cite{fidelity,formfactor,readout}.
In the present paper we show that in general we can expect a quadratic gain 
over classical computation using the algorithms of sections III-IV, and that
this gain can
be quartic for some quantities. 
The original hope
of this study was to use the quantum Fourier transform 
which is exponentially faster
than the classical Fourier transform to ensure an exponential gain
for this type of problem.  It turned out that for most systems 
counting the total number of gates involved shows
that only a polynomial gain can be reached.
However, in section V, we give an example of a related problem
where exponential gain can be reached.

\section{Semiclassical trace formula for quantum maps}

Classical and quantum maps represent a
particularly simple class of dynamical systems. 
Indeed, such systems, where one iteration of the map
corresponds to a discrete time step, are easier to handle 
and yield simpler formulas.  In what follows, we will restrict ourselves
to such systems.  This does not entail a major
loss of generality, since it is known that Hamiltonian systems
can in general be represented by maps through the construction
of Poincar\'e surfaces of section \cite{lieberman}.  Furthermore, most of
the phenomena observed in more complicated systems can be
reproduced in well-known
models of quantum maps.
This explains why many works on semiclassical formulas 
have used classical and quantum
maps as testbeds.

Here we consider two-dimensional maps on a toroidal phase space.
Let us first give examples of well-known classical maps that we will
use later on.
A much studied instance is the family of
cat maps \cite{arnold,lieberman,hannay,catkeating}, i.e.
linear automorphisms of the torus characterized by $2\times 2$ matrices
of $SL(2,\mathbb{Z})$.  For a matrix $M=\left(\begin{array}{cc}t_{11}&t_{12}\\
t_{21}&t_{22}\end{array}\right)$, the corresponding map is
\begin{eqnarray}
\bar{p} &=& t_{11} p + t_{12} q \;\;(\mbox{mod} 1) \nonumber \\
\bar{q} &= &t_{21}p + t_{22} q  \;\;(\mbox{mod} 1), 
\label{catclas}
\end{eqnarray}
where $(p,q)$ are phase-space variables
and bars denote new variables after one iteration of the map.


Another well-known example is the baker's map \cite{lieberman}:
\begin{eqnarray}
(\bar{q},\bar{p}) &= &(2q, \frac{p}{2})\;\; \mbox{for}\; 0\leq q \leq \frac{1}{2} \nonumber \\
(\bar{q},\bar{p}) &=& (2q-1, \frac{p+1}{2})\;\;\mbox{for} \; \frac{1}{2}<q\leq 1.
\label{bakclas}
\end{eqnarray}
Maps \eqref{catclas}-\eqref{bakclas} are instances of
strongly chaotic systems, with homogeneous exponential
divergence of trajectories, positive Kolmogorov-Sinai entropy, and exponential
proliferation of periodic orbits with the length.

More generally, many classical maps can be written in the form 
\begin{eqnarray}
\bar{p}& =& p - kV'(q) \nonumber \\
\bar{q}& = &q + T\bar{p},
\label{kickclas}
\end{eqnarray}
where the potential
$V(q)$ is a function of position.  Such maps
correspond to the integration over one period
of a free rotator periodically kicked by a potential $V(q)$.
They include the standard map (the 
classical version of the kicked rotator) \cite{chirikov} for $V(q)=\cos q$, 
or the sawtooth map \cite{saw1} for $V(q)=-(p-\pi)^2/2$.
  These maps display a wide range of different
behaviors depending on the parameters.  In particular, for
the standard map the dynamics changes from close to integrability
for small values of the parameter $kT$ to fully developed chaos for
large values of $kT$.
 
The quantum version of the classical maps acts on a
Hilbert space of dimension $N$ corresponding to the inverse of Planck's
constant $2\pi\hbar$. It is represented by an $N\times N$ matrix
$U$ \cite{hannay}.  
In the case of a cat map \eqref{catclas},
the quantization yields \cite{hannay,catkeating}
\begin{equation}
U_{Q_1,Q_2}=\sqrt{\frac{i t_{12}}{N}}\langle
e^{2i\pi N S(Q_1/N, Q_2/N+m)}\rangle_m,
\label{catq}
\end{equation}
where $S(q_1,q_2)=(t_{11} q_1^2-2q_1q_2+t_{22}q_2^2)/(2t_{12})$
and the average is taken over all integers $m$.

The quantized baker's map \cite{balazs} is even simpler.
The evolution operator on a $N$-dimensional space is given by
\begin{equation}
F_n^{-1}\left( \begin{array} {cc} F_{n-1}& 0 \\ 0 & F_{n-1},
\end{array} \right)
\label{bakqq}
\end{equation}
where $F_n$ is the $N\times N$ matrix 
with $(F_n)_{kj}=\frac{1}{\sqrt{N}}e^{-\frac{2i\pi kj}{N}}$ 
(discrete Fourier transform).

At last, maps of the form (\ref{kickclas}) yield, upon quantization,
quantum maps of the form:
\begin{equation}
\label{qmap}
\hat{U}=e^{-i T\hat{p}^2/2\hbar}e^{-i kV(\hat{q})/\hbar}.
\end{equation}

These evolution operators can be implemented efficiently on
a quantum computer.  This was shown for (\ref{bakqq}) in \cite{schack}
using the quantum Fourier transform
 instead
of the classical one, and in \cite{BG1,saw1} for maps of the form
(\ref{qmap}).  

One of the advantages of maps over generic systems
is that some of the steps leading to the trace formula 
linking the spectrum to periodic orbits
can be made exact. 
Indeed, the spectral density for an $N \times N$ quantum map 
$U$ with eigenphases $\theta_k$, $1\leq k\leq N$, is given by
\begin{eqnarray}
\label{density}
d(\theta)&\equiv&
\nonumber\sum_{m=-\infty}^{\infty}
\sum_{k=1}^{N}\delta\left(\theta-\theta_k+2\pi m \right)\\
&=&\frac{N}{2\pi}+\frac{1}{2\pi}\sum_{t=1}^{\infty}
\left(e^{-i t\theta}\textrm{tr}U^t+e^{i t \theta}\textrm{tr}U^{-t}\right).
\end{eqnarray}
This expression, obtained by Poisson summation formula, is exact and only 
depends on the traces of iterates of the quantum map. Similarly, one 
can express the coefficients of the characteristic polynomial
$\det\left(I-x U\right)=\sum_k\beta_k x^k$ only in terms of traces of
powers of $U$ by using the recurrence relation
\begin{equation}
\label{relation_recurrence}
\beta_k=-\frac{1}{k}\sum_{t=1}^{k}\beta_{k-t}\textrm{tr}U^t,\ \ \ \beta_0=1.
\end{equation}
This relation can be easily proved by 
expanding $\det(I+zU)=\exp$ tr $\log(I+zU)$ into powers of $z$. 
Unitarity of the operator $U$ implies the symmetry relation
\begin{equation}
\label{resurgence}
\beta_{N-k}=\det(-U)\overline{\beta_k}.
\end{equation}
Thanks to this resurgence relation the computation of tr$U^t$ for $t \leq N/2$ 
suffices to calculate the characteristic polynomial.

The semiclassical approximation of the spectrum can be obtained
by calculating the coefficients of the characteristic polynomial
\eqref{relation_recurrence} using 
semiclassical expressions for the traces. For large $N$ the main
contribution to tr$U^t$ comes from periodic orbits.
For a classical map $\phi$ mapping the phase-space onto itself,
a periodic orbit of length $t$ is a fixed point
of $\phi^t$. It is given by a sequence 
$(p_0, q_0, p_1, q_1, ..., p_t, q_t)$ of phase-space points such that
$(p_i,q_i)=\phi(p_{i-1},q_{i-1})$ for all $i, 1\leq i\leq t$, and 
$(p_t,q_t)=(p_0,q_0)$.  If $t_p$ is the smallest integer such that
$(p_{t_p},q_{t_p})=(p_0,q_0)$, then $t_p$ divides $t$ and 
the periodic orbit is the repetition of $r=t/t_p$ times a primitive
periodic orbit. A given primitive periodic orbit is characterized by its
monodromy matrix $M_p$ (which is the linearized version of the map $\phi$
in the vicinity of the periodic orbit), 
its action $S_p=\sum_{j=1}^{t_p}S(q_{i-1}, q_i)$ where
$S(q,q')$ is the classical action from $q$ to $q'$, and
its Maslov index $\nu_p$. The semiclassical expansion of tr$U^t$ reads
\begin{equation}
\label{sctrace}
\textrm{tr}U^t\approx \tau_t=\sum_{p\in\mathcal{P}_t}
\frac{t_p e^{ir( S_p/\hbar-\nu_p\pi/2)}}{|\det(I-M_p^r)|^{1/2}},
\end{equation}
where the sum runs over the set $\mathcal{P}_t$ of all periodic orbits of 
length (number of time steps) $t=r t_p$. The action, Maslov index
and monodromy matrix correspond to the associated primitive periodic orbit
 \cite{Tab}. 

For maps \eqref{catclas}-\eqref{kickclas}, the classical dynamics displays 
some form of chaos, up to the strongest types with exponential
divergence of nearby trajectories and exponential
proliferation of periodic orbits with increasing length.  Such properties
make difficult the practical use of semiclassical formulas, which need
enormous numbers of orbits to be accurate.  As we will
show in the next section, this task can be made easier 
on a quantum computer.  The fact already  mentioned that the quantum evolution 
operator of these maps can often be implemented efficiently on a
quantum computer opens the way to the use of semiclassical formulas
in the reverse direction, using quantum observables to infer
results on classical quantities.  This will be the subject of section IV.

\section{Spectrum from classical quantities}
\label{specclas}

We first discuss an algorithm allowing to obtain semiclassically
the set of eigenvalues of the quantum map, 
or equivalently the coefficients \eqref{relation_recurrence}
of the characteristic polynomial of the map.

In order to calculate the traces using \eqref{sctrace} we need to 
be able to characterize periodic orbits of the classical map. 
There are instances of systems where this task is very easy. For 
instance for cat maps \eqref{catclas}, the iterates of the 
classical map can be calculated analytically, and therefore 
periodic orbits are entirely characterized. 
This is also the case for perturbed cat maps which are Anosov maps
of the form 
$\phi=\phi_0\circ\chi_{\epsilon}$, where $\phi_0$ is a cat map and
$\chi_{\epsilon}$ is a perturbation close to the identity. It was
shown \cite{arnold88} that for sufficiently weak perturbations 
orbits of Anosov maps remain topologically conjugate to periodic orbits
of the unperturbed cat map. Thus periodic orbits can be described
completely \cite{basilio}.
More generically we will consider systems
in which periodic orbits can be described by a symbolic dynamics
associated with a finite Markov partition. That is, phase space can be
partitioned into sets $R_k$, $1\leq k\leq m$, and intersections of the
images of the $R_k$ under the (forward and backward) iterates of the
 classical map 
define finer and finer partitions so that at infinity the intersections
contain either no point or a single one \cite{AleYak81}. Thus a given 
(infinite) sequence of labels corresponds to at most one point
of phase space. The mapping rules between the 
$R_k$ under one iteration of the map can be summarized in a $m\times m$ 
transition matrix $T$ such that $T_{ij}=1$ if the image of $R_i$ has 
a non-empty intersection with $R_j$, and 0 otherwise. This transition
matrix sums up the grammar rules that discriminate between allowed words
and forbidden ones.
There is a one-to-one correspondence between phase-space 
points and allowed symbolic sequences, and periodic orbits correspond to 
periodic sequences of symbols.

Simple examples of quantum maps with symbolic dynamics are
perturbed cat maps, or the baker's map \cite{Dev86}. 
In the latter example, symbolic dynamics is described by only two symbols 
$0$ and $1$, and all sequences of symbols are allowed. 
Dynamical systems 
such as the 3-disk \cite{3disk} 
or motion on surfaces with constant negative curvature \cite{codecnc}
also provide examples where a symbolic dynamics exists with all sequences
allowed. 
In such examples, periodic trajectories are in one-to-one 
correspondence with periodic strings of $0$ and $1$.\\

We will now sketch the steps of a quantum algorithm allowing to compute the 
semiclassical traces \eqref{sctrace} in a parallel way. To simplify
notation, for each  periodic orbit $p$ of length $t=r t_p$
we define the amplitude
${\mathcal A}_p=t_p/|\det(I-M_p^r)|^{1/2}$ and the phase 
$\phi_p=r(S_p/\hbar-\nu_p\pi/2)$. Thus we have to calculate the quantities 
$\tau_t=\sum_p{\mathcal A}_p e^{i\phi_p}$.
Let us consider a system whose symbolic dynamics is described by a 
finite Markov partition. For simplicity we assume
that the partition consists of only two sets. Then only two symbols $0$ 
and $1$ are required (if there are more than two sets in the partition we
code labels by binary strings). The trace formula \eqref{sctrace} will be
truncated at $\tmax$, which means that only periodic orbits
of length $t < \tmax$ will be considered.
We distinguish five registers in the computational state. 
Register $A$ will hold the
lengths $t$, $0\leq t < \tmax$ of the periodic orbits. It
requires $\na=\log_2(\tmax)$ qubits. 
Register $B$ will hold, on its last $t$ qubits, the $2^t$
codewords corresponding to a given orbit length $t$. This register has to 
contain $\nb=\tmax$ qubits. Register $C$ is used to store the phases $\phi_p$, 
and register $D$ is used for the ''tuning'' of the amplitudes 
${\mathcal A}_p$ associated 
to each orbit. Additional registers will serve as workspace.
We will make use of following one-qubit operations: rotation
of the $k$th qubit $R_k(\theta)=\exp \left(-i\theta\sigma_y^{(k)}\right)$ 
and phase shifts $P_k(\theta)=  \exp \left(-i\theta\sigma_z^{(k)} \right)$. 
The steps are as follows.\\

{\bf Step I:} Let $\lambda$ be such that
the number of allowed codewords scales exponentially with $t$ as
$\exp(\lambda t)$. We define $\Lambda$ such that for each $t$, the 
amplitude ${\mathcal A}_p=t_p/|\det(I-M_p^r)|^{1/2}$ of each 
periodic orbit $p$ of length $t=rt_p$ is upper bounded by $\exp(-\Lambda t)$.
The amplitude ${\mathcal A}_p$ in many cases will be actually close to 
$\exp(-\Lambda t)$.
 Let us set $\mu=\Lambda-\lambda/2$.
We define angles $\theta_k \in [0,\pi/2]$ by 
$\cos\theta_k=1/\sqrt{1+e^{-2\mu 2^k}}$.
Applying $\na$ rotations $R_k(\theta_k)$ to register $A$ of the initial state
gives (up to a normalization factor) the state $\sum_{t < \tmax}\exp(-\mu t)
|t\rangle_A|0\rangle_B|0\rangle_C|0\rangle_D$. \\

\begin{figure}[hbt]
\begin{center} 
\includegraphics[width=.85\linewidth]{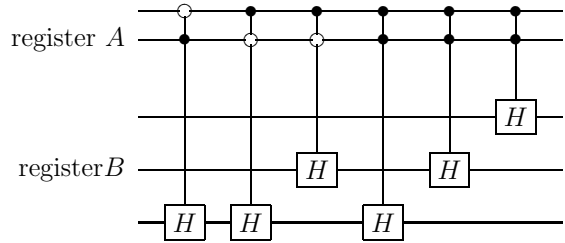}
\end{center} 
\caption{Circuit for step II and two-letter symbolic dynamics. 
Register $A$ codes for lengths $t$, $0\leq t\leq 3$ on two
qubits. The Hadamard gates are controlled by the values of $t$, 
and on register $B$ the state
$\ket{t}\ket{0}$ becomes $2^{-t/2}\sum_{i=0}^{2^t-1}\ket{t}\ket{i}$.}
\label{circuit1}
\end{figure}

{\bf Step II:} All allowed codewords are generated on register $B$. In the 
simplest case where there is no grammar rule one wants, for each value 
$t$ on register $A$, to put the last $t$ qubits of $B$ into a uniform superposition. 
This is performed by applying, for each $t$, $t$ Hadamard gates controlled 
by register 
$A$ on the last $t$ qubits of $B$ (see Fig.~\ref{circuit1}). 
This gives (up to normalization) the state
\begin{equation}
\label{etat1}
\sum_t\sum_{p}e^{-\Lambda t}|t\rangle_A|p\rangle_B
|0\rangle_C|0\rangle_D,
\end{equation}
where the second sum runs over all codewords $p$, $0\leq p\leq 2^t-1$.
If there is a finite number of grammar rules the allowed codewords 
can be generated by replacing the Hadamard gates by rotations $R_k$. 
For each value of $t$ these rotations are controlled not only by register $A$ 
(as in Fig.~\ref{circuit1}) but also by qubits of register $B$. 
Steps I and II are polynomial in $t_{max}$.\\

{\bf Step III:} From each codeword $p$ of length $t$ it is possible 
to recover the phase-space coordinates 
$(p_0, q_0, p_1, q_1, ..., p_t, q_t)$ of the trajectory coded by this 
codeword, as well as the characteristics of this trajectory: action
 $S_p$, monodromy matrix $M_p$, Maslov index $\nu_p$. These quantities
can be calculated in a parallel way by classical operations 
implemented on the quantum workspace
registers, as has been done classically in many systems
\cite{cnc,berry,gregor,charles,uzy,daniel}.
The values of $\phi_p$ and $\ln{\mathcal A}_p+\Lambda t$ 
are then calculated and written on registers $C$ and $D$. 
For the kind of systems considered here, this step is polynomial 
in $\tmax$ as the operations are performed in parallel. 
After erasing intermediate steps we get a state
\begin{equation}
\label{step3}
\sum_t\sum_{p\in\mathcal{P}_t}e^{-\Lambda t}|t\rangle_A|p\rangle_B
\ket{\phi_p}_C|\ln{\mathcal A}_p+\Lambda t
\rangle_D.
\end{equation}

{\bf Step IV:} 
As in step I we use rotations $R_k(\theta_k)$ to transfer the value 
stored in register $D$
into an exponential prefactor $\exp(\ln{\mathcal A}_p+\Lambda t)$. 
The angles $\theta_k$ are now given 
by $\cos\theta_k=1/\sqrt{1+e^{-2\kappa 2^k}}$, where the constant
$\kappa$ sets the precision that one wants to achieve on the prefactor. 
For each value of $t$ and $p$, register $D$ is the 
sum of two orthogonal components 
$|0\rangle_D$ 
and $|\psi_p \rangle_D$.
As the prefactor $e^{-\Lambda t}$  in \eqref{step3} is meant to yield a rough 
estimate of the amplitudes ${\mathcal A}_p$, it can be expected that 
the quantities $\ln{\mathcal A}_p+\Lambda t$ are small 
and that the relative 
weight of $\braket{\psi_p}{\psi_p}$ is small.

By controlled phase shifts on register $C$ the states are 
then multiplied by the 
phase factor $e^{i\phi_p}$, and step III is run backwards to 
erase register $C$. This yields
\begin{equation}
\sum_t\sum_{p\in\mathcal{P}_t} \mathcal{A}_p e^{i\phi_p}
\ket{t}_A\ket{p}_B|0\rangle_C(|0\rangle_D +|\psi_p\rangle_D)
\end{equation}

{\bf Step V:} In order to get the semiclassical traces 
$\tau_t=\sum_p{\mathcal A}_p e^{i\phi_p}$, 
we perform $\tmax$ Quantum Fourier Transforms (QFT) on register $B$.
Each QFT corresponds to a given value of $t$ and
operates on the last $t$ qubits of register $B$. That is, the gates 
of the QFT are controlled by register $A$ (as in step II, see 
Fig.~\ref{circuit1}). This yields a state
\begin{equation}
\label{qft}
\sum_t\sum_{k=0}^{2^t-1}\sum_{p\in\mathcal{P}_t}
\mathcal{A}_p e^{i\phi_p}e^{- 2i\pi k p/2^t}
\ket{t}_A|k\rangle_B|0\rangle_C(|0\rangle_D +|\psi_p\rangle_D).
\end{equation}

{\bf Step VI:}  In (\ref{qft}) the amplitude of the $|k=0\rangle_B$
 term corresponds to the semiclassical traces $\tau_t$. Therefore
we now just have to perform a quantum search of $|0\rangle_B|0\rangle_D$ in
 (\ref{qft}). This is done by amplitude amplification performed on 
registers $B$ and $D$. 
This process, which is the slowest part of our algorithm, 
requires $O(2^{\tmax/2})$ operations (as $\braket{\psi_p}{\psi_p}$
is small, the search on register $D$ is expected to contribute only
a prefactor).
It brings the state (\ref{qft})  
into a state
\begin{equation}
\sum_t\sum_{p\in\mathcal{P}_t}
\mathcal{A}_p e^{i\phi_p}
|t\rangle_A|0\rangle_B|0\rangle_C|0\rangle_D
=\sum_t\tau_t|t\rangle_A\ket{0}_B\ket{0}_C\ket{0}_D.
\end{equation}
Quantum state tomography then gives the relative values of all
semiclassical traces $\tau_t$. The knowledge of $\tau_1$ 
(easily computed classically) allows
to obtain the absolute values of the $\tau_t$, and thus the
characteristic polynomial. Because of the symmetry relation 
\eqref{resurgence} only traces up to $\tmax=N/2$ are required.
Therefore the cost of our quantum algorithm (which is essentially the
cost of amplitude amplification in step VI) is $O(2^{N/4})$.
This is to be compared with the classical cost of $O(2^{N/2})$ required
for the calculation of the semiclassical characteristic polynomial.\\

As already mentioned our algorithm aims at estimating the accuracy of 
the semiclassical approximation. Obviously, the cost of calculating 
the exact characteristic polynomial, with a scaling in $O(N^3)$, 
is far less. Thus for systems where the trace formula is exact,
the result of the semiclassical sum should only yield with much more
efforts the same result as the exact diagonalization.  For instance 
for cat maps the exact equality tr$U^t=\tau_t$ holds in Eq.~\eqref{sctrace}, 
and thus  cat maps
are not suited to studying discrepancies between 
exact and semiclassical energy levels if the full semiclassical sum is used. 
There are however 
instances of systems for which characterization of
classical periodic orbits remains easy while the traces obtained 
through the trace formula \eqref{sctrace} are truly approximations,
such as e. g. the perturbed cat maps described above.  In such cases, our 
algorithm yields the semiclassical spectrum with quadratic efficiency 
compared to classical computation. 
Besides, even 
when the trace formula is exact, its truncation is not, and therefore
its implementation has some interest 
and has been done classically in \cite{cnc,berry,charles}.
Indeed, it enables to understand the convergence properties of the
sum over periodic orbits in (\ref{density}).

\section{classical orbits from quantum operator}

Another way of estimating the accuracy of the semiclassical approximation
is to calculate how well the classical actions of the periodic
orbits are reproduced when calculated from the spectrum through
the trace formula. 
In the semiclassical approximation the trace of the 
iterates of the quantum operator can be written as a sum over periodic 
orbits. This sum can be put under the form 
$\tau_t=\sum_p \mathcal{A}_p e^{2i\pi N S_p}\approx\textrm{tr}U^t$ 
(see Eq.~\eqref{sctrace}). The actions $S_p$ calculated from
the quantum spectrum through the semiclassical formula \eqref{sctrace}
are obtained by performing a Fast Fourier Transform (FFT) on the set of traces 
$\textrm{tr}U^t$ calculated for all matrix sizes $0\leq N<\nmax $. 
The number of traces $\nmax$ to evaluate depends on the precision 
required for the actions. 

We now discuss a quantum algorithm allowing to calculate each 
trace $\textrm{tr}U^t$, for any matrix size $0\leq N<\nmax$. 
Let $m$ be the smallest integer such that $N\leq 2^m$, and $M=2^m$.
We distinguish three registers in the state vector on which the 
computation is performed. Register $A$
stores the lengths $t$ of the orbits, $0\leq t < \tmax$, on 
$\na=\log\tmax$ qubits; here $\tmax$ is some fixed integer specifying the 
highest period that we want to consider. 
The two other registers, each of length $m$,
will store the computational basis vectors. Additional workspace registers
will be used as well in the course of the computations.
Starting from the state $|0\rangle_A|0\rangle_B|0\rangle_C$, we
perform the following steps.\\

{\bf Step I:} We first apply Hadamard gates on registers $A$, $B$ and $C$ to 
put them in an equal superposition of basis vectors. We obtain
\begin{equation}
\label{step2_1}
\sum_t\sum_{i=0}^{M-1} \ket{t}_A\ket{i}_B\ket{i}_C.
\end{equation}
What we want is in fact a sum running over a range $0\leq i\leq N-1$. 
To obtain this from \eqref{step2_1}
we use an auxiliary qubit (register $D$) that is set to $|0\rangle$ if $i-N<0$ 
and to $|1\rangle$ if $i-N\geq 0$. The relative weight of
the state 
\begin{equation}
\sum_t\sum_{i=0}^{N-1}\ket{t}_A\ket{i}_B\ket{i}_C\ket{0}_D
\label{interm0}
\end{equation}
is greater than $1/2$.\\

{\bf Step II:} The $N\times N$ matrix $U^t$ has to be applied to 
register $B$ of each state $\ket{t}_A\ket{i}_B\ket{i}_C$.
As an illustration we focus on operators of the type $(\ref{qmap})$. It was
shown in \cite{BG1} that for such maps one iteration can be implemented efficiently
for a fixed matrix of size a power of 2. The algorithm consists in using 
QFTs to shift back and forth between $p$ and $q$ 
representation, while the operators $e^{i f(\hat{p})}$ and $e^{i V(\hat{q})}$ are applied
in the basis where they are diagonal by multiplication of basis vectors by a phase. 
For $N\neq 2^m$ the simulation of the quantum map involves a QFT on vectors 
of size not a power of 2. Such a procedure was proposed in \cite{kitaev} 
for any fixed vector size $N$. The simulation of $U$ can therefore be done
efficiently. 
The simulation of $U^t$ can be done sequentially, controlled by the
qubits of register $A$ (as in Fig.~\ref{circuit1}).\\

{\bf Step III:} The state \eqref{interm0} is now transformed into:
\begin{equation}
\sum_t\sum_{i}^{N-1}|t\rangle_A\left(U^t|i\rangle_B\right)|i\rangle_C\ket{0}_D
=\sum_t\sum_{i,j=0}^{N-1} U^t_{j,i}|t\rangle_A|j\rangle_B|i\rangle_C\ket{0}_D.
\end{equation}
By amplitude amplification on registers $B$, $C$, $D$ we select vectors 
with $\ket{j}_B=\ket{i}_C$ and $\ket{0}_D$, leading to
\begin{equation}
\sum_t\sum_{i=0}^{N-1} U^t_{i,i}|t\rangle_A|i\rangle_B|i\rangle_C \ket{0}_D.
\end{equation}
After erasing register $C$ we perform a QFT on register $B$. 
As in section \ref{specclas}, we use amplitude amplification to select the 
state $\ket{0}_B$, whose amplitude is 
$\sum_i U^t_{i,i}/\sqrt{M}=\textrm{tr}U^t/\sqrt{M}$. 
This is the slowest step in our computation.
For chaotic systems the matrix elements $U^t_{i,j}$ for $N\times N$ matrices
are of order $1/\sqrt{N}$ and the traces $\textrm{tr}U^t$ are expected 
to be of order 1. Thus each amplitude amplification has a cost 
$O(\sqrt{N})$ and step III requires $N$ Grover iterations in total. 
For integrable systems the traces are of order $\sqrt{N}$, and therefore 
only one of the amplitude amplifications is needed, requiring
$\sqrt{N}$ Grover iterations in total for step III.\\

{\bf Step IV:} We are now in the state
\begin{equation}
\sum_t\textrm{tr}U^t|t\rangle_A\ket{0}_B\ket{0}_C\ket{0}_D.
\end{equation}
The relative values of the traces for different values of $t$ are obtained by 
quantum state tomography. The traces themselves are then deduced from the 
classical calculation of tr$U$, requiring $O(N)$ classical operations.\\

The algorithm requires the calculation 
of $\nmax$ traces $\textrm{tr}U^t$, with $0\leq N<\nmax $.
Thus the cost of the quantum algorithm is of order $\nmax^2$.
Classically, we need to compute all $\nmax$ traces. Except for 
tr$U_N$ this would need $O(N^2)$ classical operations if the map is
of the type $(\ref{qmap})$, and up to $O(N^3)$ in the general case where 
diagonalization of the operator is required. Thus the classical cost 
is of order $\nmax^3$ to $\nmax^4$ operations. Thus in both cases
the quantum algorithm outperforms classical computation, 
albeit polynomially.

We note that if one is interested in distinguishing 
integrable and chaotic systems via the form factor as in the
algorithm proposed in \cite{formfactor}, then one needs only
to be able to distinguish traces of order $\sqrt{N}$ (integrable case)
from ones of order $1$ (chaotic case), for a specific value of $N$.
In this case one can stop at step III and check that 
$\sqrt{N}$ Grover iterations are enough to get to the state $\ket{0}$, 
in which case one concludes that the system is integrable, 
or not enough, in which case  one concludes that the system is
chaotic. Our algorithm then only needs 
$O(\sqrt{N})$ quantum operations instead of $O(N^2)$ classical operations, 
an improvement from the quadratic gain in \cite{formfactor}.  One
can also compute exactly the trace (stopping at step IV),
and compute the form factor for small $t$, with a quadratic improvement
compared to classical computation.

\section{Exponential speed-up by phase estimation}
\label{expcat}

The preceding processes can be applied to many physical systems and yield a
polynomial speed-up compared to classical computation. However there exist 
systems where a larger (up to exponential) gain might be obtained, following
a different strategy based on phase estimation. This method 
\cite{kitaev,ablloyd}
enables to obtain an eigenvalue of a given operator $U$ 
by applying conditionally
iterates of $U$ to an eigenvector $\ket{\Psi}$; 
this gives the state $\sum_i|i\rangle U^i|\Psi\rangle$
which, by Fourier transform on the first register,
 gives $|\theta\rangle|\Psi\rangle$,
where $\exp(i\theta)$ is the eigenvalue corresponding
to $\ket{\Psi}$.
If $\ket{\Psi}$ is not an eigenvector but some randomly chosen state,
the same process leads to $\sum_j \alpha_j|\theta_j\rangle|\Psi_j\rangle$
where $|\Psi_j\rangle$ are eigenvectors and 
$\ket{\Psi}=\sum_j \alpha_j |\Psi_j\rangle$.
To be efficient, this method critically requires not only that 
$U$ can be efficiently implemented, but also that exponential iterates 
of $U$ can be implemented with polynomial number of quantum gates,
a much more stringent requirement.  In the case of the quantum
cat map, this method is efficient and remarkably enough
can lead to classical quantities with exponential efficiency.

It is known \cite{hannay} that the $n$th iterate of the quantized
cat map (\ref{catq}) coincides with the quantization of the classical 
$n$th iterate.
In \cite{catqc}, it was shown that one can simulate
the classical cat map efficiently on a quantum computer, while in
\cite{BGpo}, it was further 
shown that one can compute the classical $n$th iterate
for exponentially large $n$ with polynomial number of gates. 
Thus if one starts from a random vector $\ket{\Psi}$, one can compute
 $\sum_i|i\rangle U^i|\Psi\rangle$ in polynomial number 
of gates for exponential $i$'s and $N$; a quantum Fourier 
transform followed by a
quantum measurement leads to the value $|\theta_j \rangle$ of one eigenvalue of
the quantum cat map.  It is known \cite{catkeating} 
that these eigenvalues are very
constrained, being of the form 
\begin{equation}
\theta_j=\frac{2\pi j + \phi(N)}{n(N)},
\end{equation}
where $n(N)$ is the quantum period function, that is 
the smallest integer such that
\begin{equation}
U^{n(N)}=I e^{i\phi(N)},
\end{equation}
and the phase $\phi(N)$ can be calculated easily 
from the components of matrix $L$ \cite{catkeating}. 

Thus using this algorithm the quantum period function can be obtained 
in polynomial time on a quantum computer.  This quantity is related to the
classical period function, which for each matrix $L$ is the 
shortest integer $g$ such that $L^g=I \; \textrm{mod} N$. 
Indeed, the quantum period
$n(N)$ is also the smallest integer such that $L^{n(N)}=I\; (\textrm{mod} N)$
if $N$ is odd, and such that 
$L^{n(N)}=\left( \begin{array} {cc} 1 \;(\textrm{mod} N)& 0\; (\textrm{mod} 2N) 
\\ 0 \;(\textrm{mod} 2N) & 1 \; (\textrm{mod} N) \end{array} \right)$ if $N$ 
is even.  
The two functions in all cases differ by at most a factor of two 
\cite{catkeating}, so knowing one of them enables to test and find easily the
other one. The classical 
period function describes the periodic orbits of the classical cat map.
It has been shown in \cite{BGpo} that finding it is as complex
as factorization of integers, and can nevertheless be realized on a quantum
computer polynomially fast using a variant
of order-finding.  The use of the quantum cat map
enables to get this classical quantity by an equally efficient alternate 
quantum algorithm, showing that in this specific case classical
quantities can be obtained through quantum mechanics with exponential
efficiency compare to classical algorithms.

\section{Conclusion}

In the studies above, we have shown that it is possible to implement
semiclassical formulas on quantum computers, with a gain on efficiency 
over the implementation on a classical computer.  The gain is in
general polynomial, but in specific instances an exponential gain 
can be obtained for related problems.  We mention 
again that the algorithms of Section IV can also be used to study
 quantum systems without reference to the semiclassical approximation,
in the manner of \cite{formfactor}, with actually a larger gain.

The algorithms of
sections III and IV can be generalized to a large class of systems.
Indeed, to generalize section III
one can use the tool of Poincar\'e surface of section to 
transform systems with continuous time to discrete maps.
 For example, a popular system to study quantum chaos corresponds to
billiards, i.e. classically a particle bouncing between walls, and quantum mechanically a wave function obeying Helmholtz equation with boundary conditions.
In this case, a simple surface of section is represented by the boundary,
the phase space coordinates being the curvilinear abscissa along the boundary
and the angle that the outgoing trajectory makes with the vector
normal to the boundary.  An alternate possibility would be to stick
with the continuous time dynamics and use the semiclassical formulas
appropriate for this case.  In both cases, it is important to be able
to enumerate the classical trajectories used in the semiclassical sums,
which requires that a reasonably good symbolic dynamics can be constructed
(e.g. with finite Markov partition).  This is already the case for classical
implementations of these formulas, which have all been performed in such cases.
Additionally, the method exposed in 
section III is all the more efficient since the Lyapunov exponent of
orbits is uniform.  In the case where the stability of different orbits
varies widely in different phase space regions, the quantum algorithm will become
less efficient.  Thus although strongly chaotic systems are the
most difficult to treat by semiclassical formulas, they are probably the ones 
where the algorithms above will be the most efficient compared to classical
algorithms.

To generalize Section IV to systems with continuous time is probably possible,
but would necessitate to first build an algorithm to simulate such
systems on quantum computers.  We think that once this is done,
the main ideas of our algorithm in section IV should then be applicable.

The quantum algorithms presented here can be applied to a wide variety
of systems.  They show that in a domain where extensive
numerical simulations have been used in the past decades, a quantum
computer could significantly improve the speed of the calculations.

We thank the French ANR
(project INFOSYSQQ) and the IST-FET program of the EC
(project EUROSQIP) for funding.


\vskip -0.5cm
\begin{thebibliography}{99}
\bibitem{feynman} R.~P.~Feynman, Found. Phys. {\bf 16}, 507 (1986)
\bibitem{josza} A.~Eckert and R.~Josza, Rev. Mod. Phys. {\bf 68}, 733 (1996).
\bibitem{steane} A. Steane, Rep. Progr. Phys. {\bf 61}, 117 (1998).
\bibitem{nielsen} M. A. Nielsen and I. L. Chuang, {\em Quantum computation
                  and quantum information}, Cambridge Univ. Press, 2000.
\bibitem{shor} P.~W.~Shor, in Proc. 35th Annu. Symp. Foundations of
               Computer Science (ed. Goldwasser, S. ), 124 
               (IEEE Computer Society, Los Alamitos, CA, 1994).
\bibitem{grover} L.~K.~Grover, Phys. Rev. Lett. {\bf 79}, 325 (1997).
\bibitem{lloyd} S.~Lloyd, Science {\bf 273}, 1073 (1996).
\bibitem{ablloyd} D.~S.~Abrams and
                S.~Lloyd, Phys. Rev. Lett. {\bf 79}, 2586 (1997).
\bibitem{BG1}B.~Georgeot and D.~L.~ Shepelyansky, Phys. Rev. Lett.
                 {\bf 86}, 2890 (2001).
\bibitem{saw1} G.~Benenti, G.~Casati, S.~Montangero and D.~L.~Shepelyansky, 
                 Phys. Rev. Lett. {\bf 87}, 227901 (2001).
\bibitem{interm} O.~Giraud and B.~Georgeot, 
Phys. Rev. A {\bf 72}, 042312 (2005). 
\bibitem{einstein} A. Einstein. Verh. Dtsch. Phys. Ges. {\bf 19}, 82 (1917).
\bibitem{gutz}  M. C. Gutzwiller, J. Math. Phys. (N.Y.) {\bf 12}, 343 (1971); 
R. Balian and C. Bloch, Ann. Phys. (N.Y.) {\bf 85}, 514 (1974).
\bibitem{vanvleck} J. H. Van Vleck, Proc. Natl. Acad. Sci. USA {\bf 14}, 
           178 (1928).
\bibitem{miller}  W. H. Miller, Adv. Chem. Phys. {\bf 25}, 69 (1974).
\bibitem{cnc} R.~Aurich, M.~Sieber and F.~Steiner,
 Phys. Rev. Lett. {\bf 61}, 483 (1988).
\bibitem{berry} M.~V.~Berry, Nonlinearity {\bf 1}, 399 (1988).
\bibitem{gregor}G. Tanner, P. Scherer, E. B. Bogomolny, B. Eckhardt, and D. Wintgen, Phys. Rev. Lett. {\bf 67}, 2410 (1991).
\bibitem{charles} E. Bogomolny and C. Schmit, Nonlinearity {\bf 6}, 523 (1993).
\bibitem{uzy} H.~Primack and U.~Smilansky, J. Phys. A: Math. Gen. {\bf 31},
             6253 (1998).
\bibitem{daniel} D.~Braun, P.~A.~Braun and F.~Haake, Physica D {\bf 131}, 265
           (1999); D.~Braun, Chaos {\bf 9}, 730 (1999).
\bibitem{curvature} P. Cvitanovic and B. Eckhardt, Phys. Rev. Lett. {\bf 63}, 823 (1989). 
\bibitem{berrykeating}  M. V. Berry and J. P. Keating, 
Proc. Rev. Soc. London A {\bf 437}, 151 (1992).
\bibitem{harmonic} J.~Main, V.~A.~Mandelshtam and H.~S.~Taylor. 
Phys. Rev. Lett. {\bf 79}, 825 (1997). 
\bibitem{fidelity}J. Emerson, 
Y. S. Weinstein, S. Lloyd and D. Cory,
Phys. Rev. Lett. {\bf 89}, 284102 (2002).
\bibitem{formfactor} D. Poulin, R. Laflamme, G. J. Milburn and J. P. Paz,
Phys. Rev. A {\bf 68}, 022302 (2003).
\bibitem{readout} B.~Levi, B.~Georgeot and D.~L.~Shepelyansky, Phys. Rev. E 
{\bf 67}, 046220 (2003);
M.Terraneo, B.Georgeot and D.L.Shepelyansky, 
Phys. Rev. E {\bf 71}, 066215 (2005).
\bibitem{lieberman}A.~Lichtenberg and M.~Lieberman, {\em Regular and Chaotic 
                  Dynamics}, Springer, N.Y. (1992).
\bibitem{arnold}V.~I.~Arnold and A.~Avez, {\em Ergodic Problems of Classical 
               Mechanics}, Benjamin, N. Y. (1968).
\bibitem{hannay} J.~H.~Hannay and M.~V.~Berry, Physica D {\bf 1}, 267 (1980).
\bibitem{catkeating} J.~Keating,  Nonlinearity {\bf 4}, 277 (1991); 
{\bf 4}, 309 (1991).
\bibitem{chirikov} B.~V.~Chirikov, Phys. Rep. {\bf 52}, 263 (1979).
\bibitem{balazs} N. L. Balazs and A. Voros, Ann. Phys. (N.Y.) {\bf 190}, 
           1 (1989). 
\bibitem{schack} R.~Schack, Phys. Rev. A {\bf 57}, 1634 (1998).
\bibitem{Tab} M. Tabor, Physica D {\bf 6}, 195 (1983).
\bibitem{arnold88}V.~I.~Arnold, {\em Geometrical methods in the theory of differential equations}, Springer (1988).
\bibitem{basilio} M.~Basilio de Matos and A.~M.~Ozorio de Almeida,
                  Ann. Phys. {\bf 237}, 46 (1995).
\bibitem{AleYak81}V.~M.~Alekseev and M~.V.~Yakobson, Phys. Rep. {\bf 75}, 
290 (1981).
\bibitem{Dev86}R.~L.~Devaney, {\em An Introduction to 
Chaotic Dynamical systems}, Benjamen, Menlo Park, 1986.
\bibitem{3disk}  P.~Cvitanovic, B.~Eckhardt, P.~E.~Rosenqvist, 
      G.~Russberg and P.~Scherer, 
in G. Casati and B. Chirikov, eds., {\em Quantum Chaos}, 
        (Cambridge University Press, Cambridge 1994).
\bibitem{codecnc} C.~Series, J. London Math. Soc. {\bf s2-31}, 69 (1985).
\bibitem{kitaev} A.~Kitaev, Electronic Colloquium on 
Computational Complexity (ECCC)
{\bf 3} nr 3 (1996) (also preprint quant-ph/9511026).

\bibitem{catqc} B.~Georgeot and D.~L.~ Shepelyansky, Phys. Rev. Lett. 
              {\bf 86}, 5393 (2001).
\bibitem{BGpo} B.~Georgeot, Phys. Rev. A {\bf 69}, 032301 (2004).
\end{thebibliography}
\end{document}